\documentclass[11pt]{article}
\begin{document}
\title{{\bf $\kappa$-Minkowski and Snyder algebra from 
reparametrisation symmetry }}
\author{
{\bf Chandrasekhar Chatterjee $^{}$\thanks{chandra@bose.res.in}}, 
{\bf Sunandan Gangopadhyay $^{ }$\thanks{sunandan@bose.res.in} }\\
 S.~N.~Bose National Centre for Basic Sciences,\\JD Block, Sector III, 
Salt Lake, Kolkata-700098, India\\[0.3cm]
}
\date{}

\maketitle

\begin{abstract}
\noindent Following our earlier work \cite{sunandan1, sunandan2}, we derive noncommuting phase-space structures which are combinations of both the 
$\kappa$-Minkowski and Snyder algebra
by exploiting the reparametrisation symmetry of the recently
proposed Lagrangian for a point particle \cite{subir} satisfying the exact
Doubly Special Relativity dispersion relation in the
Magueijo-Smolin framework.
\\[0.3cm]
\\[0.3cm]

\end{abstract}

\noindent Introduction :\\

\noindent Recently, motivated by the ideas of quantum gravity 
\cite{smolin}, a generalization of Special Relativity known as
Doubly Special Relativity (DSR) \cite{amelino} has been proposed. 
\noindent The most popular model known as Magueijo-Smolin (MS)
DSR \cite{smo1,smo2} has the dispersion relation
\begin{eqnarray}
\phi_1=p^{2}-m^{2}\left[1-\frac{\eta p}{\kappa}\right]^2 \approx 0
\label{3}
\end{eqnarray}
where, $\eta^{0}=1$, $\vec{\eta}=0$.
As one can see that there is an additional mass scale $\kappa$
which plays an important role in two very different contexts.
The observations of ultra-high energy cosmis ray particles
and photons that violate the GZK bound \cite{piran} can be explained
by using the deformed dispersion relation (\ref{3}). On the other hand,
the existence of a length scale is directly linked to the breakdown of
spacetime continuum and the emergence of a noncommutative (NC) spacetime
\cite{rob,sny}. Furthermore, exploiting the notion of duality in the 
context of Quantum Groups, it has been demonstrated in
\cite{luk,now} that each DSR is uniquely associated with a particular
form of NC phase-space. In particular, the MS relation is related to a 
specific representation of $\kappa$-Minkowski NC phase-space 
\cite{luk,now}.

\noindent So far the theoretical development in this area has been mainly
kinematical. A dynamical picture of this model has been proposed
recently in \cite{subir}. It is shown that the reparametrisation
symmetry of the proposed Lagrangian allows one to
choose appropriate gauge fixing conditions such that 
specific forms of NC phase-space structures are induced via Dirac brackets
\cite{dirac}. A further change of variables is made which leads to
an algebra which is a combination of both the 
$\kappa$-Minkowski and Snyder.
However, there does not seem to be a systematic prescription
on which such change of variables are derived.

\noindent In this paper (essentially following the method in 
\cite{sunandan1,sunandan2}) we exploit the reparametrisation
symmetry of this Lagrangian to derive systematically
a change of variables from which a NC phase-space structure 
containing both the 
$\kappa$-Minkowski \cite{amelino, luk, now} 
and the Snyder algebra \cite{sny} emerges.\\

\noindent DSR Lagrangian and NC phase-space :\\

\noindent We start from the reparametrisation invariant action of
an MS particle \cite{subir}\footnote{$x^{\mu}$ are the 
spacetime coordinates,
the dot here denotes
differentiation with respect to the evolution parameter $\tau$,
and $g_{\mu\nu}$ represents the flat Minkowski metric with
$g_{00}=-g_{ii}=1$.}
\begin{eqnarray}
S &=& \int d\tau\left[\frac{m\kappa}{\sqrt{\kappa^2-m^2}}
\left(g_{\mu\nu}\dot{x}^{\mu}\dot{x}^{\nu}+
\frac{m^2}{\kappa^2-m^2}(g_{\mu\nu}\dot{x}^{\mu}\eta^{\nu})^2
\right)^{1/2} - \frac{m^2 \kappa}{\kappa^2-m^2}g_{\mu\nu}
\dot{x}^{\mu}\eta^{\nu}\right]\nonumber\\
&\equiv&\int d\tau\left[\frac{m\kappa}{\sqrt{\kappa^2-m^2}}
\Lambda-\frac{m^2 \kappa}{\kappa^2-m^2}(\dot{x}\eta)\right]
\label{1}
\end{eqnarray}
where, $\Lambda=\left(g_{\mu\nu}\dot{x}^{\mu}\dot{x}^{\nu}+
\frac{m^2}{\kappa^2-m^2}(g_{\mu\nu}\dot{x}^{\mu}\eta^{\nu})^2
\right)^{1/2}$.

\noindent The canonically conjugate momenta to $x^{\mu}$ are given by
\begin{eqnarray}
p_{\mu}&=&\frac{m\kappa}{\sqrt{\kappa^2-m^2}}
\frac{(\dot{x}_{\mu}+\frac{m^2}{\kappa^2-m^2}(\dot{x}\eta)\eta_{\mu})}
{\Lambda}-\frac{m^2 \kappa}{\kappa^2-m^2}\eta_{\mu}~.
\label{2}
\end{eqnarray}
They are subject to the DSR dispersion (constraint) relation (\ref{3})
and satisfy the Poisson bracket (PB) relations
\begin{eqnarray}
\{x_{\mu},p_{\nu}\}=-g_{\mu\nu}\quad;\quad \{x_{\mu}, x_{\nu}\}
 = \{p_{\mu}, p_{\nu}\} = 0~.
\label{4}
\end{eqnarray}
Now using the reparametrisation symmetry of the problem 
(under which the action (\ref{1}) is invariant)
and the fact that $x^{\mu}(\tau)$ transforms as a  scalar
under world-line reparametrisation
\begin{eqnarray}
\tau \rightarrow \tau' &=& \tau'(\tau)\nonumber\\
x^{\mu}(\tau)\rightarrow x'^{\mu}(\tau') &=& x^{\mu}(\tau)
\label{5}
\end{eqnarray}
leads to the following infinitesimal transformation of the space-time
coordinate
\begin{eqnarray}
\delta x^{\mu}(\tau) = x'^{\mu}(\tau) - x^{\mu}(\tau)
=\epsilon \frac{dx^\mu}{d\tau}~. 
\label{6}
\end{eqnarray}
The generator of this reparametrisation invariance is obtained by 
first writing the variation in the 
Lagrangian $L$ (\ref{1}) under the 
transformation (\ref{6}) as a total derivative
\begin{eqnarray}
\delta L = \frac{dB}{d\tau} \quad;\quad 
B=\frac{\epsilon A\kappa\Lambda}{1+A^2}\left(1-\frac{\eta p}{\kappa}\right)
\label{totalder}
\end{eqnarray}
where, $A=\frac{m}{\sqrt{\kappa^2-m^2}}$~.
Now the generator $G$ is obtained from the usual Noether's prescription as
\begin{eqnarray}
G=\frac{1}{2}\left(p^{\mu}\delta x_{\mu} - B\right) 
= \frac{\epsilon\Lambda}{2A\kappa}\phi_{1}
\label{7}
\end{eqnarray}
where we have used (\ref{6}, \ref{totalder}). 
It is easy to see that this generator reproduces 
the appropriate transformation (\ref{6})
\begin{eqnarray}
\delta x^{\mu}(\tau)=\{x^{\mu}, G\}=\epsilon \frac{dx^\mu}{d\tau}~. 
\label{8}
\end{eqnarray}
The simplest gauge condition to get rid of the gauge freedom generated
by $\phi_{1}$ (\ref{6}) is obtained by 
identifying the time coordinate $x^{0}$ with the evolution parameter $\tau$ 
\begin{equation}
\phi_{2} = x^{0} - \tau \approx 0
\label{9.1}.
\end{equation}
The constraints (\ref{3}, \ref{9.1}) form a second class set with
\begin{equation}
\{\phi_{a}, \phi_{b}\} = 2m^2
\left(\frac{p_{0}}{A^2\kappa^2}+\frac{1}{\kappa}\right)
\epsilon_{ab}\quad;\quad (a, b=1,2).
\label{9.2}
\end{equation}
The resulting non-vanishing Dirac brackets (DB) are\footnote{The
Dirac brackets are defined as 
$\{A, B\}_{DB} = 
\{A, B\} - \{A, \phi_{a}\}(\phi^{-1})_{ab}\{\phi_{b}, B\}$, 
where $A$, $B$ are any pair 
of phase-space variables \cite{dirac}.}:
\begin{eqnarray}
\{x_{i}, p_{0}\}_{DB} = \frac{p_{i}}{(\frac{p_{0}m^2}{A^2\kappa^2}+
\frac{m^2}{\kappa})}\qquad; 
\qquad\{x_{i}, p_{j}\}_{DB} = -g_{ij}
\label{9.3}
\end{eqnarray}
which imposes the constraints $\phi_{1}$ and $\phi_{2}$ strongly.

\noindent Using (\ref{8}), the transformations that relates
the primed coordinates in terms
of the unprimed coordinates can be written down in terms
of phase-space variables as:
\begin{eqnarray}
x'^{0} &=& x^{0} + \epsilon \nonumber\\
x'^{i} &=& x^{i} + \frac{\epsilon(1+A^2)}{(p_{0}+A^2\kappa)}p^{i} 
\label{10}
\end{eqnarray}
where we have used the relation 
$\frac{dx^{i}}{d\tau}=\frac{(1+A^2)}{(p_{0}+A^{2}\kappa)}p^{i}$.
The above change of variables (derived from reparametrisation
symmetry) enables us to choose a value 
of the reparametrisation parameter $\epsilon$ which 
leads to noncommuting structures which are combinations 
of both the $\kappa$-Minkowski and the Snyder algebra as we shall see subsequently.

\noindent Setting
\begin{eqnarray}
\epsilon=-\frac{(p_{0}+A^2 \kappa)}{A^2 \kappa^2(1+A^2)}(x.p)
\label{value}
\end{eqnarray}
and using (\ref{9.3}) and (\ref{10}), we obtain the following algebra
between the primed coordinates :
\begin{eqnarray}
\{x'_{i}, p_{j}\}&=&-g_{ij} + \frac{\kappa^2-m^2}{\kappa^{2}m^2}
p_{i}p_{j}
\label{101aa}
\end{eqnarray}
\begin{eqnarray}
\{x'_{i}, x'_{j}\}&=&\frac{\kappa^2-m^2}{\kappa^{2}m^2}
(x'_{i}p_{j}-x'_{j}p_{i})
\label{alg1}
\end{eqnarray}
\begin{eqnarray}
\{x'_{0}, x'_{i}\}&=&\frac{1}{(1+A^2)}
\left[-\frac{1}{\kappa}x'_{i}-\frac{1}{A^{2}\kappa^2}x'_{i}p_{0}\right]
+\frac{1}{A^{2}\kappa^2}x'_{0}p_{i}
\label{alg2}
\end{eqnarray}
\begin{eqnarray}
\{x'_{i}, p_{0}\}&=&\frac{1}{\kappa}p_{i}
+\frac{\kappa^2-m^2}{\kappa^{2}m^2}
p_{i}p_{0}
\label{alg3}
\end{eqnarray}
\begin{eqnarray}
\{x'_{0}, p_{i}\}&=&\frac{1}{\kappa}p_{i}
+\frac{\kappa^2-m^2}{\kappa^{2}m^2}
p_{i}p_{0}~.
\label{alg4}
\end{eqnarray}
Now making the following change of variables
\begin{eqnarray}
\tilde{x}'_{\mu}&=&M_{\mu\nu}x'^{\nu} 
\label{alg5}
\end{eqnarray}
where,
\begin{equation}
M_{00}=1+A^2\quad,\quad 
M_{0i}=M_{i0}=0,\quad M_{ij}=g_{ij}
\label{alg5a}
\end{equation}
we obtain the algebra which contains both the
$\kappa$-Minkowski \cite{amelino, luk, now} and the Snyder \cite{sny} NC structures :
\begin{eqnarray}
\{\tilde{x}'_{\mu}, \tilde{x}'_{\nu}\}&=&
\frac{1}{\kappa}(\tilde{x}'_{\mu}\eta_{\nu}-\tilde{x}'_{\nu}\eta_{\mu})
+\frac{\kappa^2-m^2}{\kappa^{2}m^2}
(\tilde{x}'_{\mu}p_{\nu}-\tilde{x}'_{\nu}p_{\mu})
\label{alg6}
\end{eqnarray}
\begin{eqnarray}
\{\tilde{x}'_{\mu}, p_{\nu}\}&=&
-g_{\mu\nu}+
\frac{1}{\kappa}(p_{\mu}\eta_{\nu}+p_{\nu}\eta_{\mu})
+\frac{\kappa^2-m^2}{\kappa^{2}m^2}
p_{\mu}p_{\nu}
\label{alg7}
\end{eqnarray}
\begin{eqnarray}
\{p_{\mu}, p_{\nu}\}=0~.
\label{alg7a}
\end{eqnarray}
It should be noted that in absence of 
the $1/\kappa$-term or the 
$(\kappa^2-m^2)/(\kappa^2 m^2)$-term, one obtains
the Snyder \cite{sny} or the $\kappa$-Minkowski algebra
\cite{amelino, luk, now} respectively.

\noindent Alternatively, one may demand that the algebra
between $x'_{i}$ and $p_{j}$ is of the form
(\ref{101aa}). A simple inspection (after the substitution
of (\ref{10}) in the left hand side of (\ref{101aa})) gives
the solution (\ref{value}) for the reparametrisation parameter
$\epsilon$.

\noindent Hence, the change of variables relating the primed coordinates with
the unprimed ones read:
\begin{eqnarray}
x'_{0} &=& x_{0} - \frac{\kappa^2-m^2}{\kappa^{2}m^2}
\frac{(p_0+A^{2}\kappa)}{(1+A^2)}(x.p)\nonumber\\
x'_{i} &=& x_{i} - \frac{\kappa^2-m^2}{\kappa^{2}m^2}(x.p)p_{i}~. 
\label{10aa}
\end{eqnarray}
With the above change of variables and the DB algebra (\ref{9.3}),
one can reproduce the algebra between the primed variables
(\ref{alg1}, \ref{alg2}, \ref{alg3}, \ref{alg4}). Note that
the above change of variables is different 
from the one given in \cite{subir}.

\noindent Furthermore, the solution for the reparametrisation
parameter (\ref{value}) shows that one can write down a modified
gauge fixing condition given by :
\begin{eqnarray}
\phi_3=x^0+\frac{(p_{0}+A^2 \kappa)}{A^2 \kappa^2(1+A^2)}(x.p)
-\tau\approx0
\label{modifiedgauge}
\end{eqnarray}
which forms a second class pair with (\ref{3}).
The set of non-vanishing DB(s) consistent with the strong imposition
of the constraints (\ref{3}, \ref{modifiedgauge}) reproduces the results
(\ref{101aa}, \ref{alg1}, 
\ref{alg2}, \ref{alg3}, \ref{alg4}).

\noindent At this point we would like to make an 
observation. It can be easily seen by substituting
(\ref{10}) in (\ref{alg3}) and (\ref{alg4}) (by keeping 
only the first terms on the right hand side of
(\ref{alg3}) and (\ref{alg4}) which are the $\kappa$-Minkowski terms) that there
exists no solution of the reparametrisation parameter 
$\epsilon$ which simultaneously
satisfy both (\ref{alg3}) and (\ref{alg4}). This in turn
shows that it is not possible to obtain 
the $\kappa$-Minkowski algebra by exploiting the
reparametrisation symmetry of the action (\ref{1}) for the
MS particle. Hence we point out that although
the action (\ref{1}) (for the MS particle) proposed in
\cite{subir} gives the DSR dispersion relation
(\ref{3}), it does not lead directly to the algebra
of the $\kappa$-Minkowski space rather it leads to
an algebra which contains both the $\kappa$-Minkowski
and the Snyder algebra. This is also consistent
with the fact that the analysis in \cite{subir} also does
not reproduce the algebra of the 
$\kappa$-Minkowski space.\\

\noindent Discussions : \\
In this paper, we have essentially followed our earlier work on
noncommutativity and reparametrisation symmetry \cite{sunandan1, sunandan2}
to obtain phase-space NC structures which are combinations
of both the $\kappa$-Minkowski and Snyder algebra from the 
recently proposed Lagrangian for a point particle  
satisfying the exact DSR dispersion relation.
We have also shown that 
the NC results in the nonstandard gauge
and the commutative results in the standard gauge are seen
to be gauge transforms of each other.
We feel our approach is conceptually cleaner
and more elegant than \cite{subir} where such change 
of variables are found by inspection and 
apparently lack any connection with the 
inherent symmetries of the problem.\\

\noindent Acknowledgement: The authors are thankful
to the referee for making useful comments.



\begin{thebibliography}{99}
\bibitem{smolin}C. Rovelli, L. Smolin, Nucl. Phys. B. 442 (1995) 593,
erratum: ibid 456 (1995) 734.
\bibitem{amelino}G. Amelino-Camelia, Nature 418, (2002) 34;
Phys. Lett. B. 510, (2001) 255. 
\bibitem{smo1}J. Magueijo, L. Smolin, Phys. Rev. Lett. 88 (2002) 190403.
\bibitem{smo2}J. Magueijo, L. Smolin, Phys. Rev. D. 67 (2003) 044017;
[gr-qc/0207085].
\bibitem{piran}G. Amelino-Camelia, T. Piran, Phys. Rev. D. 64 (2001) 036005;
[astro-ph/0008107].
\bibitem{rob}S. Doplicher, K. Fredenhagen, J.E. Roberts, 
Phys. Lett. B. 331 (1994) 39.
\bibitem{sny}H.S. Snyder, Phys. Rev. 71 (1947) 68.
\bibitem{luk}J. Lukierski, A. Nowicki, H. Ruegg, V.N. Tolstoy,
Phys. Lett. B. 264 (1991) 331.
\bibitem{now}J. Kowalski-Glikman, S. Nowak, Phys. Lett. B. 539 (2002) 126;
[hep-th/0203040].
\bibitem{subir}S. Ghosh, Phys. Rev. D. 74 (2006) 084019;
[hep-th/0608206].
\bibitem{sunandan1}R. Banerjee, B. Chakraborty, S. Gangopadhyay; 
J. Phys. A. 38, 957 (2005), [hep-th/0405178].
\bibitem{sunandan2}S. Gangopadhyay, 
J. Math. Phys. 48, (2007) 052302, [hep-th/0703174].
\bibitem{dirac}P.~A.~M.~Dirac,
{\it Lectures on Quantum Mechanics,}
Belfer Graduate School of Science, Yeshiva University,
New York, 1964.






\end{thebibliography}
\end{document}